\documentclass[12pt]{article}
\usepackage{graphicx}

\begin{document}

\begin{center}
\Large {\bf Entanglement Entropy in Quantum Gravity and \\the Plateau Problem}
\end{center}

\bigskip
\bigskip

\begin{center}
Dmitri V. Fursaev
\end{center}

\bigskip
\bigskip

\begin{center}
{\it Dubna International University \\
  and\\
  the University Centre,\\
  Joint Institute for Nuclear Research\\
  141 980, Dubna, Moscow Region, Russia\\}
 \medskip
\end{center}
\begin{center}
  {\rm E-mail: \texttt{fursaev@theor.jinr.ru}}\\
  \medskip
\end{center}

\bigskip
\bigskip

\begin{abstract}
In a quantum gravity theory the entropy of entanglement $S$ between the
fundamental degrees of freedom spatially divided by a surface is discussed.
The classical gravity is considered as an emergent phenomenon
and arguments are presented that: 1) $S$
is a macroscopical quantity which can be
determined  without knowing a real microscopical content of the fundamental theory; 2)
$S$ is given by the Bekenstein-Hawking formula in terms of the area of a
co-dimension 2 hypesurface $\cal B$;  3) in static space-times $\cal B$
can be defined as a minimal hypersurface of a least volume separating the system
in a constant time slice.
It is shown that properties of $S$ are in agreement with
basic properties of the von Neumann entropy. Explicit variational formulae for $S$
in different physical examples are considered.
\end{abstract}

\newpage

\section{Introduction}

The concept of gravity as an emergent phenomenon \cite{Sakh} continues to be
one of the attracting and stimulating ideas in theoretical physics
\cite{Vol}, \cite{Padm}.
According to this
point of view, the space-time metric is an effective low-energy variable and metric changes
are collective excitations of underlying degrees of freedom of some unknown fundamental theory.
The idea gets an inspiration from the condensed matter
physics. For example, at very small scales the space-time may have a discrete lattice
structure and gravitons, like phonons, may be collective excitations of the lattice.
Gravity analogs in condensed matter physics may be very useful to develop an intuition about
the physics at Planckian energies and to model phenomena predicted in quantum gravity \cite{Vol}.

The experience with condensed matter models tells us that
macroscopical phenomena can be described without knowledge of the microscopical structure.
Under certain conditions one can quantize the effective low-energy
variables and get  reliable results which are equivalent to computations
in terms of genuine microscopical degrees of freedom \cite{VoZe}. This feature
is also present in effective field theories, like the theory of pions (see
\cite{Burgess} and references therein).

One of characteristics of a  many-body system is a quantum entanglement which measures
a degree of the  correlation between different parts
of the system. Quantum entanglement in condensed matter received a considerable
attention in last few years. Its study helps to understand  better collective effects
in strongly correlated systems, see \cite{AFOV} for a review.
Another application of the entanglement is a quantum information theory where it
is used to store and process the information \cite{Pres}.

The quantum entanglement is quantified by an  entropy.
The present paper considers the entanglement entropy in a fundamental
(quantum gravity) theory
by adopting the point of view that gravity is an emergent phenomenon \cite{Sakh}.

Let us start with the problem formulated in \cite{DF:06a}: if the fundamental
theory were known what would be the
entropy of entanglement in a flat space-time between the degrees of freedom
divided by a plane?
According to \cite{DF:06a} in the leading order approximation the entropy $S$
for a sufficiently large but finite piece $\cal B$ of the plane can be
expressed as
\begin{equation}\label{i.1}
S={{\cal A}({\cal B}) \over 4G}~~~,
\end{equation}
where ${\cal A}$ is the area of $\cal B$.
It follows from (\ref{i.1}) that the main
information about correlations of the degrees of freedom across the plane is encoded in the
gravity coupling $G$ in the low-energy sector of the theory.
The conjecture was motivated by the black hole
physics where the Bekenstein-Hawking entropy $S^{BH}$ is given by (\ref{i.1})
and ${\cal A}$ is the area of the black hole horizon.
One of possible interpretations \cite{Sr:93}--\cite{FrNo:93}
relates $S^{BH}$  to an entanglement of the degrees of freedom across
the  horizon.

The aim of the present paper
is to find a definition of the entanglement entropy $S$ in the presence of the gravitational
field and study its properties.
The case considered in \cite{DF:06a} is a system divided by a plane.
In a flat space the plane  is a minimal surface.
We present a number of arguments that the entanglement entropy
in a quantum gravity  is given by  (\ref{i.1}) and
in a static space-time the separating surface $\cal B$
is a minimal least area hypersurface in a constant-time slice.

Finding a minimal surface for given boundary conditions is called the Plateau problem.
The fact that $S$ is a dynamical variable which has a simple geometrical
meaning is rather non-trivial from the physical point of view. It means that
{\it the entanglement entropy  in a quantum gravity theory can be measured
solely in terms of
macroscopical (low-energy) parameters without the knowledge of a microscopical
content of the theory}. In this regard $S$, being a more general concept, is analogous
to a thermodynamical entropy.

The relation between the entanglement entropy and minimal surfaces is not
unexpected. In a different context, Ryu and Takayanagi \cite{RT:06a}, \cite{RT:06b}
suggested a holographic formula
for the entanglement entropy in conformal field theories (CFT) which admit a dual description
in terms of the anti-de Sitter (AdS) gravity one dimension higher. The holographic formula
looks as (\ref{i.1}) where $G$ is a higher-dimensional gravity coupling. It yields the entropy
of a CFT in terms of the area of a minimal surface embedded in a higher-dimensional
bulk space-time. The holographic formula
can be tested in case of two-dimensional CFT's and it reproduces correctly results
of quantum computations.

Although our hypothesis and the suggestion of  \cite{RT:06a}, \cite{RT:06b} may be related
they are different
statements. The holographic formula is a geometrical
representation
of the entanglement entropy for a certain class of quantum field models. Thus, the
entropy here is
a model and cutoff dependent quantity.

The paper organized as follows. The arguments in favour of (\ref{i.1}) are presented in section 2.
We remind a path integral approach for computation of the entanglement entropy in QFT's
and introduce  a  "partition function" $\cal Z$ in a form of a Gibbons-Hawking
path integral with a special choice of the boundary conditions. The entropy $S$ can be
computed from $\cal Z$ in a statistical-mechanical manner.
A similar construction was used in \cite{DF:06b} to justify the
holographic formula of \cite{RT:06a}, \cite{RT:06b}.
Quantum fluctuations of the geometry do not allow
to fix the separating surface at the beginning. The surface is specified by
boundary conditions. We show that
the leading order approximation to $\cal Z$ is determined by the geometries where the separating
surface is a minimal co-dimension 2 hypersurface with a least volume.

The main suggestion about entanglement entropy in a quantum gravity theory
is formulated in section 3 for static space-times.
We first consider space-times where constant-time slices are Cauchy hypersurfaces
with simply connected boundaries. Static space-times are singled out because
the separating surface $\cal B$, which is an extremal co-dimension 2 hypersurface
in the space-time, is a minimal
co-dimension 1 hypersurface in a constant time slice. The latter feature
ensures that $S$ has the properties of the von Neumann entropy.

The Cauchy surface of a maximally extended
Schwarzschild black hole space-time has two disconnected asymptotically
flat regions. In section 4 we use this example to formulate the suggestion
about the entanglement entropy
in the
presence of a black hole horizon. The bifurcation surface of the horizon is a
minimal surface on a constant-time slice.
One can relate to this surface an entanglement entropy which
coincides with the Bekenstein-Hawking entropy and
corresponds to the loss of information about states located behind the horizon.

The arguments of section 2 based on the path integral approach serve as a motivation.
They cannot be considered as a real proof of (\ref{i.1}) because they concern
an effective theory.
In section 5,
we show that our suggestions regarding the entanglement
entropy have an important feature:
they are consistent with the subadditivity property of the von Neumann entropy.
The material of this section is similar to the
analysis of the entropy in CFT's with  gravity duals \cite{HiTa:06}, \cite{HeTa:07} .

We give examples how to express changes of $S$ solely in terms
of macroscopical parameters of the system in section 6. The examples include changes
caused by shifts of a
point-like particle and by deformations of a cosmic string located near a separating surface.
In section 7 we discuss quantum corrections to formula
(\ref{i.1}) from low-energy fields and suggest that ultraviolet divergences of an entanglement
entropy encountered in QFT models are removed in the course of the standard renormalization
of gravity couplings. The paper is ended with a discussion of its results
in section 8.

\section{Formulation of the problem}
\setcounter{equation}0

\subsection{Entanglement entropy in QFT}

We first recall a path-integral approach to calculation
of entanglement entropy in QFT models \cite{DF:06a}.
Consider a stationary space-time $\cal M$ with
the number of dimensions $d$ (the concrete value of $d$ will not be important
for our analysis). Let $\partial_t$ be a time-like Killing vector field of $\cal M$.
Suppose that the system is defined on a domain $\Sigma$ of
a constant time slice $t=\mbox{const}$.

We also assume that the system is in a thermal state with the temperature $T$. The corresponding
density matrix is $\hat{\rho}={\cal N}^{-1}\exp(-\hat{H}/T)$ where $\hat{H}$ is the Hamiltonian
which is defined as a generator of the evolution of the given system along the Killing time $t$.

\begin{figure}[h]
\begin{center}
\includegraphics[height=4.5cm,width=6.0cm]{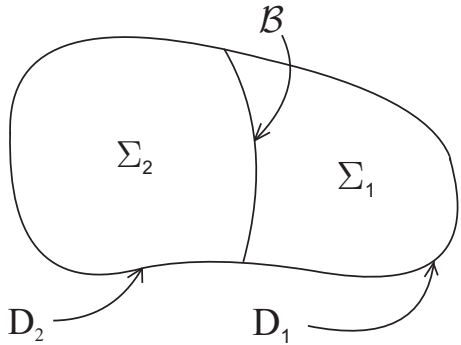}
\caption{\small{A constant time slice of a system spatially divided by a
hypersurface $\cal B$.}}
\label{f1}
\end{center}
\end{figure}

\begin{figure}[h]
\begin{center}
\includegraphics[height=5.5cm,width=14.0cm]{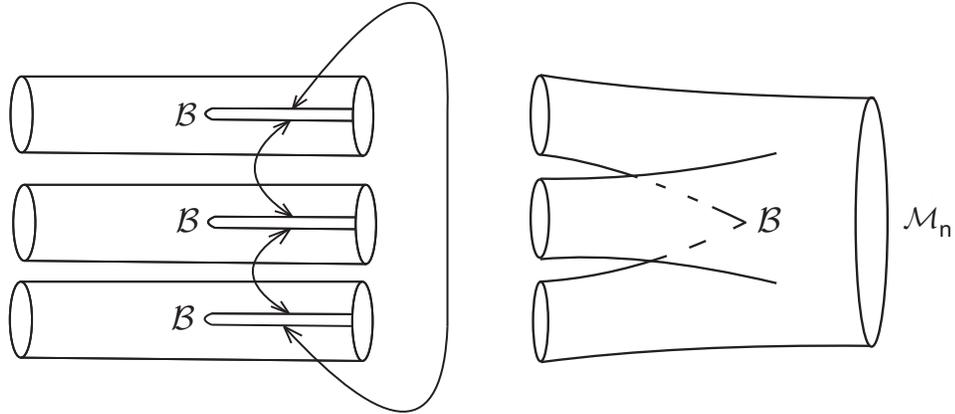}
\caption{\small{Construction of the space ${\cal M}_n$  for a
2D finite-temperature system set on an interval $\Sigma$. For $n=3$
the space ${\cal M}_n$ is obtained from 3 copies of the cylinders $\Sigma \times S^1$.
The cylinders are cut and glued
along the part $\Sigma_1$ of a "constant time" slice.}}
\label{f2}
\end{center}
\end{figure}

Suppose that $\Sigma$ is divided into two parts, $\Sigma_1$ and $\Sigma_2$, by a
co-dimension 2 hypersurface $\cal B$ (see Fig. \ref{f1}) and define a reduced density matrix
for the region $\Sigma_1$,
\begin{equation}\label{rdm}
\hat{{\rho}}_1=\mbox{Tr}_2 \hat{\rho}~~,
\end{equation}
where the trace is taken over the states located in the region $\Sigma_2$.
The entanglement entropy is defined as
\begin{equation}\label{2.3}
S_1=-\mbox{Tr}_1 \hat{{\rho}}_1\ln \hat{{\rho}}_1~~.
\end{equation}
Analogously, one can define the entanglement entropy $S_2$ in terms of the reduced density matrix
obtained by tracing the degrees of freedom in $\Sigma_1$.
There is a method of computation of $S_1$ based on the relation
\begin{equation}\label{2.3a}
S_1=-\lim_{n\rightarrow 1}~{\partial \over \partial n}\mbox{Tr}_1
~\hat{{\rho}}_1^{~n} ~~.
\end{equation}
Consider the partition function of the given system
$$
Z(T)=\mbox{Tr}~e^{-\hat{H}/T}~~~.
$$
According to the standard approach it can be written in terms of the path integral
\begin{equation}\label{2.4a}
Z(T)=\int [D\phi]~e^{-I[\phi,T]}~~~,
\end{equation}
where $I[\phi,T]$ is the classical action for the given model. The fields
$\phi$ in (\ref{2.4a}) are defined on
a Riemannian manifold ${\cal M}=\Sigma\times S^1$ where $S^1$ is a circle of the length $T^{-1}$.

The recipe for computing the entropy $S_1$ is the following \cite{DF:06a}.
Consider a family of manifolds ${\cal M}_n$ where $n$ is a natural number and
${\cal M}_1\equiv {\cal M}$. To obtain ${\cal M}_n$ for $n>1$
one has to cut ${\cal M}$ along one of its constant time hypersurfaces  isomorphic
to $\Sigma$. The cut is made along the domain $\Sigma_1$ of $\Sigma$.
Then one takes $n$ identical copies of the manifolds  and
glues them along the cuts. This procedure is shown  on Fig. \ref{f2} for a 2D theory.
As a next step, for each $n$ one defines a "partition function" on ${\cal M}_n$
\begin{equation}\label{2.4}
Z_1(\beta,T)=\int [D\phi]~e^{-I[\phi,\beta,T]}~~~,
\end{equation}
as a direct generalization of (\ref{2.4a}).
The parameter $\beta$ in (\ref{2.4}) is $2\pi n$. The entropy
can be represented in a statistical-mechanical form
\begin{equation}\label{2.5}
S_1(T)=-\lim_{\beta \rightarrow 2\pi}~ \left(\beta {\partial \over
\partial \beta}-1\right) \ln Z_1(\beta,T)~~~.
\end{equation}
The operation with $\beta$ in (\ref{2.5}) should be understood in the following way:
one first computes $Z_1(\beta,T)$ for $\beta=2\pi n$, and then replaces
$\beta$ with a continuous parameter. This can be done
even if ${\cal M}_n$ itself cannot be defined at arbitrary $\beta$.
Equation (\ref{2.5}) coincides with the formula for the
entropy in statistical mechanics if $\beta^{-1}$
is interpreted  as a
temperature. Note that $\beta^{-1}$ is a geometrical parameter, the physical
temperature of the system is $T$. By cutting $n$ copies of ${\cal M}$ and gluing along $\Sigma_2$ one can obtain
the entanglement entropy in the region $\Sigma_2$.

The reduced density matrix (\ref{rdm})
is given by a functional integral like (\ref{2.4a}) on $\cal M$ with a cut along $\Sigma_1$.
The values of the field variables on the both sides of the cut determine the "indexes"
of the matrix $\hat{{\rho}}_1$. Thus, the trace like
$\mbox{Tr}_1~\hat{{\rho}}_1^{~n}$ is determined by the path integral (\ref{2.4}) and
(\ref{2.5}) follows from (\ref{2.3a}).

The method described above enables one to compute the entanglement entropy in quantum
field models. The calculation can be done in the regularized theory
in the presence of an ultraviolet cutoff. The dependence on the cutoff $\varrho$ has a very
simple form \cite{Sr:93}, \cite{BKLS}. For instance,
the leading divergence in the entropy in a four-dimensional
theory is proportional to $A/\varrho^2$, where $A$ is the area of the separating surface
$\cal B$ ($\varrho$ is assumed to have the dimension of a length).
In condensed matter systems the entanglement entropy is finite and $\varrho$ is a
physical parameter associated, for example, with a lattice spacing.

\subsection{Basic assumptions}

To define the entanglement entropy in a quantum gravity theory one has to answer
a number of questions. Here are some of them.

1) The notion of classical geometry may not hold at microscopical scales.
Does the definition of a "separating surface" make sense in a quantum gravity theory?

2) Related to the first question is the problem of the entanglement
of gravitational degrees of freedom. How  can one take into account
graviton contributions to the entropy?

3) The entanglement entropy of quantum fields is divergent. Can
this problem be solved by a traditional prescription based on the renormalization?
What are the physical constants which should be renormalized?

To approach these problems we assume that the gravity along with other low-energy fields
(the fields of the Standard Model, for example) is an emergent phenomenon, in a sense that
the fields
represent an effective description of some other underlying fundamental variables.
One can guess that it is the effective nature of the low-energy theory which results
in ultraviolet divergences, while the underlying theory does not have this problem.
We assume that the trace over the fundamental degrees of freedom which has to be done
to compute the reduced density
matrix in the quantum gravity is equivalent to the trace over low-energy variables
including fluctuations of the metric.

In the present paper we restrict our analysis by theories at finite-temperatures, including
vacuum states as a limiting case.
This yields
an opportunity to use Euclidean formulation of the theory.
By following Gibbons and Hawking we work with a partition function for a
canonical ensemble in a gravity theory \cite{GH2}, \cite{Hawk:79}
\begin{equation}\label{2.1}
Z[h,\varphi]=\int [Dg][D\phi]\exp(-I[g,\phi])~~~,
\end{equation}
where $I[g,\phi]$ is a classical Euclidean action for the metric $g_{\mu\nu}$ and
some set of matter fields $\phi$.
The path integral (\ref{2.1}) is considered
as a low-energy approximation for a genuine partition function of a quantum gravity theory.

The "trajectories"  in (\ref{2.1}) are
Riemannian manifolds $\cal M$ and all possible field configurations living on them.
The number of dimensions of $\cal M$ is $d$, $d>2$.
Two dimensional gravity models can be included in our analysis but they require a separate
discussion.

The form
of "trajectories" is specified by the boundary conditions.
It is assumed that manifolds
have a boundary
$\partial {\cal M}={\cal T}={\cal S} \times S^1$
where ${\cal S} $ is a $(d-2)$-dimensional spatial boundary of the system.
We demand that $\cal T$ possesses a Killing vector field $\xi=\partial_\tau$ with
closed orbits $S^1$.
The length of the circle $S^1$ is fixed by the inverse temperature.
It will be assumed that ${\cal S}$ is a compact closed manifold.
A typical example of the boundary is ${\cal S}=S^2$.
The partition function depends on the metric $h_{\mu\nu}$ of ${\cal T}$ and on
values $\varphi$ of matter fields on ${\cal T}$.

\begin{figure}[h]
\begin{center}
\includegraphics[height=6.1cm,width=8.0cm]{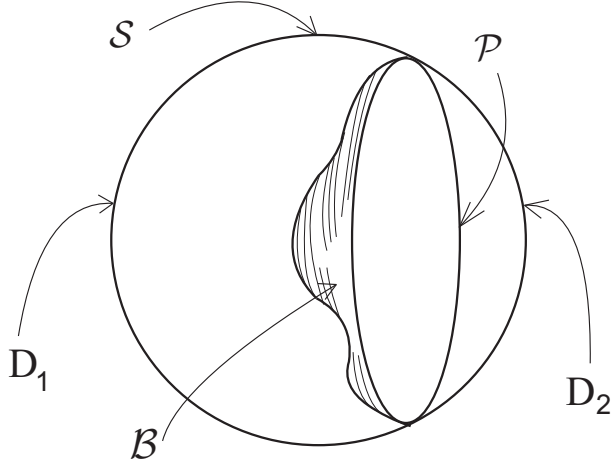}
\caption{\small{In a quantum gravity one can specify the entangled regions of a system
by dividing its spatial boundary
${\cal S}$  onto domains $D_1$ and $D_2$.
The separating surface $\cal B$ is determined by a dynamical problem with the condition that
$\cal B$ ends at the boundary $\cal P$ between $D_1$ and $D_2$.}}
\label{f3}
\end{center}
\end{figure}

The action in (\ref{2.1}) is
\begin{equation}\label{2.2a}
I[g,\phi]=I[g]+I_{\mbox{m}}[g,\phi]~~,
\end{equation}
\begin{equation}\label{2.2}
I[g]=-{1 \over 16\pi G}\int_{\cal M} d^{d}x\sqrt{g}
\left(R-2\Lambda\right)
-{1 \over 8\pi G}\int_{\partial \cal M} d^{d-1}x\sqrt{h}K~~.
\end{equation}
Here $I_{\mbox{m}}[g,\phi]$ is an action of matter fields on $\cal M$ whose explicit
form is not relevant for further discussion.
The boundary term in (\ref{2.2}) is the Gibbons-Hawking term which
depends on the trace of the extrinsic curvature $K$ of
${\cal T}$ embedded in $\cal M$. This term is needed to formulate
a well-defined variational procedure.

In the semiclassical approximation the partition function takes the form
\begin{equation}\label{2.10a}
Z[h,\varphi]\sim \exp(-I[\bar{g}(h),\bar{\phi}(\varphi)])~~~,
\end{equation}
where $\bar{g}(h)$ and $\bar{\phi}(\varphi)$ are solutions to classical
equations which define extrema of  $I[g,\phi]$ for given boundary
conditions.
Equations for the
metric are the (Euclidean) Einstein equations
\begin{equation}\label{2.10}
R_{\mu\nu}-\frac 12 \bar{g}_{\mu\nu} R+\Lambda \bar{g}_{\mu\nu}=8\pi G T_{\mu\nu}~~~,
\end{equation}
with a matter stress-energy tensor  $T_{\mu\nu}$. We denote by $\bar{\cal M}$
a manifold whose metric is a solution to (\ref{2.10}). It is assumed that
the boundary condition $\partial \bar{\cal M} ={\cal T}$ admits stationary
solutions.

Because in a quantum gravity theory the metric except its
boundary value cannot be fixed a priory,
for the same reason the definition of
entangled states of a system can be described only in terms of a partition of the boundary.
We take a constant time ($\tau=const$) slice $\cal S$ of $\cal T$ and
consider division of $\cal S$ by a hypersurface ${\cal P}$, see Fig. \ref{f3}.
The parts $D_1$ and $D_2$ of $\cal S$ separated by ${\cal P}$
will be treated as boundaries
of two entangled regions.

Let  $S(D_1,T)$ be an entanglement entropy which appears when the degrees
of freedom inside the region with the boundary $D_2$ are traced out.
By bearing in mind the method described in the previous section we
relate the entropy to a partition function
\begin{equation}\label{2.51}
S(D_1,T)=-\lim_{\beta \rightarrow 2\pi}~ \left(\beta {\partial \over
\partial \beta}-1\right) \ln {\cal Z}(\beta,D_1,T)~~~.
\end{equation}
The function ${\cal Z}(\beta,D_1,T)$ is determined by a
parameter $\beta=2\pi n$ where $n$ is a natural number.  We assume that
${\cal Z}(\beta,D_1,T)$ can be represented in terms of a sum over geometries, ${\cal M}_n$,
specified by a boundary condition $\partial {\cal M}_n={\cal T}_n$.

The boundary space ${\cal T}_n$ is constructed from ${\cal T}$ as follows.
One cuts ${\cal T}$ along the part $D_2$ of the slice $\tau=const$ and glues
$n$ identical copies of ${\cal T}$ along the cuts. Spaces
${\cal T}_n$ have conical singularities on a surface where the cuts meet.
The singular surface is isometric to
$\cal P$. The structure of ${\cal T}_n$ near the singular surface
is ${\cal C}_\beta \times {\cal P}$ where
${\cal C}_\beta$ is a 2-dimensional cone with the circumference length $\beta=2\pi n$ around
the tip.

For $n>1$ ($\beta>2\pi$) we define the partition function as:
\begin{equation}\label{2.1a}
{\cal Z}(\beta,D_1,T)=\sum_{\cal B}Z(\beta,D_1,{\cal B},T)~~,
\end{equation}
\begin{equation}\label{2.1aa}
Z(\beta,D_1,{\cal B},T)=\int' [Dg][D\phi]\exp(-I[g,\phi])~~~.
\end{equation}
The path integral $\int'$ in the right hand side of (\ref{2.1aa}) is similar to (\ref{2.1})
but
it is taken over a class of special geometries,
$\{{\cal M}_n\}_{\cal B}$  (and field configurations living on them).
The boundary of spaces from $\{{\cal M}_n\}_{\cal B}$
is ${\cal T}_n$.
Because ${\cal T}_n$ have conical singularities $\{{\cal M}_n\}_{\cal B}$ are also
supposed to have conical singularities on a co-dimension 2
hypersurface $\cal B$.
The surface $\cal B$ has the following properties:  it is located
entirely in ${\cal M}_n$ and intersects the boundary
${\cal T}_n$ at $\cal P$.
Near $\cal B$ the spaces look as ${\cal C}_\beta \times {\cal B}$.
Inside the class $\{{\cal M}_n\}_{\cal B}$ the metric of $\cal B$ is fixed.

To summarize,  calculation of ${\cal Z}(\beta,D_1,T)$ consists of two steps. One
takes the path integral (\ref{2.1aa}) over the class of spaces specified by the
presence of the singular surface $\cal B$ and then one sums
contributions of different classes $\{{\cal M}_n\}_{\cal B}$. As a result,
all possible geometries which obey the boundary conditions and have conical
singularities on co-dimension 2 hypersurfaces are taken into account.
The definition of ${\cal Z}(\beta,D_1,T)$ is motivated by the
form of the QFT partition function (\ref{2.4}).

The summation symbol
$\sum_{\cal B}$ in (\ref{2.1a}) does not have a precise meaning,
the situation common for path integrals.
This is not a principle difficulty because we are dealing with an effective
theory and our main result will be based on a semiclassical approximation.

\subsection{Semiclassical approximation}

To compute the integral (\ref{2.1aa}) in the semiclassical approximation
one has to take into account the form of the classical gravity action
on   $\{{\cal M}_n\}_{\cal B}$,
\begin{equation}\label{2.7}
I[g,\beta]=-{1 \over 16\pi G}\int_{{\cal M}_n} d^{d}x\sqrt{g}
\left(R-2\Lambda\right)
-{1 \over 8\pi G}\int_{\partial {\cal M}_n} d^{d-1}x\sqrt{h}K
-{1 \over 8\pi G}(2\pi-\beta){\cal A}({\cal B})~~.
\end{equation}
Here ${\cal A}({\cal B})$ is the volume of ${\cal B}$.
The first term in the r.h.s. of (\ref{2.7}) is an integral over a regular domain
of ${\cal M}_n$, the last term appears as a result of
conical singularities on $\cal B$ \cite{FS1}\footnote{Conical singularities
of the boundary space ${\cal T}_n$
 do not contribute to (\ref{2.7}). This can be
shown by using arguments presented in \cite{DF:06b}.}.

The semiclassical approximation to (\ref{2.1aa}) is determined by  extrema
of the action (\ref{2.7}) on $\{{\cal M}_n\}_{\cal B}$.
Because the area of  $\cal B$ is fixed metric variations
should vanish at $\cal B$. Thus, extrema of (\ref{2.7}) are geometries
which are solutions to gravity equations (\ref{2.10}) outside $\cal B$.

Let $\bar{\cal M}_n$ be such a solution
for a class $\{{\cal M}_n\}_{\cal B}$. Any solution $\bar{\cal M}_n$ with $n>1$ can
be constructed from a solution
$\bar{\cal M}$ with $n=1$.
Consider to this aim an embedding of $\cal B$ in $\bar{\cal M}$,
such that $\cal B$ ends on a constant time slice of the boundary $\cal T$.
One cuts $\bar{\cal M}$  along a co-dimension 1 hypersurface
which starts at the domain $D_2$ of the slice on $\cal T$
and ends on $\cal B$ (the precise choice of the surface does not matter).
Spaces $\bar{\cal M}_n$ are obtained by gluing $n$ identical copies
with the same cut so that to get conical singularities on $\cal B$. Because the cut intersects the slice of
$\cal T$ along $D_2$ the constructed spaces also satisfy the required boundary conditions,
$\partial \bar{\cal M}_n={\cal T}_n$.

In the semiclassical approximation
\begin{equation}\label{2.8a}
-\ln Z(\beta,D_1,{\cal B},T)\simeq
I[\bar{g}(h),\beta]+
I_{\mbox{m}}[\bar{g}(h),\bar{\phi}(\varphi),\beta]=
n I[\bar{g}(h),\bar{\phi}(\varphi)]+
\mu_\beta ~{\cal A}({\cal B})~~~,
\end{equation}
where  $\mu_\beta\equiv (n-1)/(4G)=(\beta -2\pi)/(8\pi G)$ and
$\bar{g}(h)$, $\bar{\phi}(\varphi)$ are corresponding classical solutions.
The gravity action $I[\bar{g}(h)]$ is given by (\ref{2.7}). The total action
$I[\bar{g}(h),\bar{\phi}(\varphi)]$ in the r.h.s. of (\ref{2.8a}) is set
on
$\bar{\cal M}$ and does not depend on $\cal B$.
The
partition function (\ref{2.1a}) can be written in the form
\begin{equation}\label{2.9}
{\cal Z}(\beta,D_1,T)\sim e^{-nI[\bar{g}(h),\bar{\phi}(\varphi)]}
\sum_{\cal B}e^{-\mu_\beta {\cal A}({\cal B})}~~.
\end{equation}
The sum $\sum_{\cal B}$ goes over surfaces embedable in $\bar {\cal M}$.
Because $\mu_\beta>0$ the sum  is dominated by the contribution from a surface $\bar{\cal B}$
which  has a least area in $\bar {\cal M}$ for given boundary conditions. Therefore,
\begin{equation}\label{2.9a}
-\ln{\cal Z}(\beta,D_1,T)\simeq
n I[\bar{g}(h),\bar{\phi}(\varphi)]+\mu_\beta~
{\cal A}(\bar{\cal B})~~.
\end{equation}
By using (\ref{2.51}) and (\ref{2.9a}) one gets the entanglement entropy
in a quantum gravity theory for the region with boundary $D_1$
\begin{equation}\label{2.12}
S(D_1,T)\simeq {{\cal A}(\bar{\cal B}) \over 4G}~~~.
\end{equation}
The entropy
is given in terms of the area of a surface separating the
two regions with boundaries $D_1$ and $D_2$.
The formula is in agreement with the conjecture of \cite{DF:06a}.
As follows from (\ref{2.12}), at the semiclassical level the entropy for
the region with the boundary $D_2$
has the same form, $S(D_2,T)\simeq S(D_1,T)$.

It is the presence of gravitational degrees of freedom in the functional integral (\ref{2.1aa})
which results in (\ref{2.12}).
The reason why the entropy appears in the semiclassical approximation
is explained by the fact that metric is an effective low-energy variable. One needs to know
genuine degrees of freedom of the quantum gravity
to write the entropy in a von Neumann form (\ref{2.3}).

As technical remark, note that the partition function ${\cal Z}(\beta,D_1,T)$
does not coincide with
the Gibbons-Hawking path integral (\ref{2.1}) in the limit $\beta=2\pi$.
The both functionals
at the semiclassical level are determined by the same solution $\bar{\cal M}$
to the Einstein equations. However, they differ by a multiplier ${\cal N}$
present in
${\cal Z}(\beta,D_1,T)$. Formally ${\cal N}$ counts a "number" of
hypersurfaces  embedable in $\bar{\cal M}$ under given boundary
conditions. This difference is not essential because one can remove ${\cal N}$ by
changing normalization of ${\cal Z}(\beta,D_1,T)$.

\subsection{The Plateau problem}

It is convenient to fix the position of $\cal B$
with respect to some coordinate chart, $X^\mu$, $\mu=1,..,d$, on $\bar{\cal M}$. The embedding
can be locally described in the parametric form
\begin{equation}\label{3.1}
X^\mu=X^\mu(y^i)~~~,
\end{equation}
where $y^i$, $i=1,..,n=d-2$ are some coordinates on $\cal B$.
If $\bar{g}_{\mu\nu}$ is the metric on $\bar{\cal M}$ the metric induced on $\cal B$ is
$\gamma_{ij}=\bar{g}_{\mu\nu} X^\mu_{,i}X^\nu_{,j}$. The volume of $\cal B$ is
\begin{equation}\label{2.11aa}
{\cal A}({\cal B})=\int_{\cal B} d^ny \sqrt{\gamma} ~~~,
\end{equation}
where $\gamma=\det \gamma_{ij}$.
Because hypersurfaces embedded in $\bar{\cal M}$  differ by their position
we look for local extrema of the volume functional with respect to the variation
of coordinates $X^\mu$ under the fixed metric $\bar{g}_{\mu\nu}$ and fixed boundary,
\begin{equation}\label{2.11}
\delta_X{\cal A}({\cal B})=0~~~.
\end{equation}
Hypersurfaces with such a property are called {\it minimal}  \cite{Fomenko}.
The entropy (\ref{2.12}) is determined by a  least volume minimal hypersurface, i.e. by
a {\it global
minimum} of the volume functional. Finding a minimal surface
for a given boundary is called the {\it Plateau problem}.
If the background metric $\bar{g}_{\mu\nu}$ has a Lorentzian signature the hypersurfaces
which obey condition
(\ref{2.11}) are called {\it extremal}.

The r.h.s. of (\ref{2.11}) can be rewritten as
\begin{equation}\label{2.11a}
\delta_X{\cal A}({\cal B})=\int_{\cal B} d^ny \sqrt{\gamma}
\gamma^{ij}X^\mu_{~,i} X^\nu_{~,i}~\delta X_{\mu;\nu} ~~~,
\end{equation}
where
$\delta X_{\mu;\nu}=\bar{g}_{\mu\lambda}\nabla_\nu \delta X^{\lambda}$
and $\nabla_\mu$ are the connections
defined with respect to  $\bar{g}_{\mu\nu}$. If $l^\mu$ is a unit vector
orthogonal
to the boundary $\partial {\cal B}$ and tangent to  $\cal B$
the boundary conditions imply that $\delta X^\mu l_\mu=0$ on $\partial {\cal B}$.
From (\ref{2.11}) and (\ref{2.11a}) one gets a
Euclidean version of the Nambu-Goto equations
\begin{equation}\label{3.2}
\triangle X^\mu+\Gamma^\mu_{\lambda\nu} X^\lambda_{,i}X^\nu_{,j}\gamma^{ij}=0~~~,
\end{equation}
where $\triangle$ is a Laplace operator on $\cal B$ and $\Gamma^\mu_{\lambda\nu}$
are the Christoffel coefficients for $\bar{g}_{\mu\nu}$.

The conditions for the minimal surface
can be also written in another form. Let $\delta X_{\bot}^\mu$ be a part of
the vector $\delta X^{\mu}$ which is orthogonal to $\cal B$. One can write
$\delta X_{\bot}^{\mu}=\alpha n^\mu+\beta p^\mu$, where $\alpha=
n^\mu\delta X_{\mu}$, $\beta=p^\mu\delta X_{\mu}$ and
$n^\mu$, $p^\mu$ are two unit mutually orthogonal normals to $\cal B$. One finds
\begin{equation}\label{3.3}
\delta_X{\cal A}({\cal B})=\int_{\cal B} d^ny \sqrt{\gamma}
\gamma^{ij}X^\mu_{~,i} X^\nu_{~,i}~ \delta X_{{\bot}\mu;\nu}=
\int_{\cal B} d^ny \sqrt{\gamma}(\alpha k_n+\beta k_p)~~~,
\end{equation}
where $k_n=\gamma^{\mu\nu}\nabla_\mu n_\nu$ and $k_p=\gamma^{\mu\nu}\nabla_\mu p_\nu$ are
traces of extrinsic curvatures of $\cal B$ defined with the help of
the projector $\gamma_{\mu\nu}=g_{\mu\nu}-n_\mu n_\nu-p_\mu p_\nu$.
The contribution in (\ref{3.3}) of the longitudinal part of the variation
$\delta X_{\|}^{\mu}$ (tangent to $\cal B$) is a
total divergence which vanishes due to the boundary conditions.
Therefore, equations (\ref{2.11a}) are equivalent to
\begin{equation}\label{3.4}
k_n= k_p=0~~.
\end{equation}

\begin{figure}[h]
\begin{center}
\includegraphics[height=4.3cm,width=14.0cm]{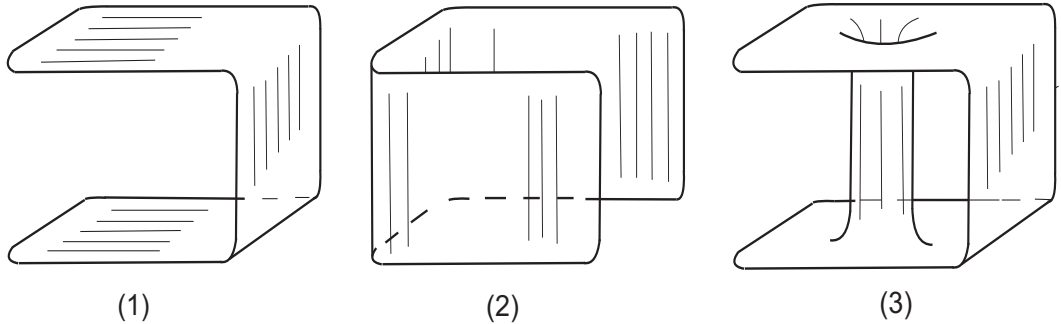}
\caption{\small{Some examples of minimal surfaces corresponding to a single contour
in a flat space.}}
\label{f7}
\end{center}
\end{figure}

We complete this section with brief comments regarding the Plateau problem \cite{Fomenko}.
The theory of minimal surfaces appeared from consideration of soap films
and it was known since 18-th century. The Plateau problem does not have a
unique solution. Even in a flat space there may exist more than one minimal surface
for a given boundary contour, see Fig. \ref{f7} (1,2,3). The solution is unique if the contour lies in a plane and
does not have self-intersections. Note that in three-dimensions
there may be contours which cannot be boundaries of minimal
surfaces with a finite area. Such contours and their higher dimensional
analogs are excluded from our analysis.

In general minimal surfaces can be represented as a combination of parts of smooth
surfaces glued along their boundaries. These boundaries may appear as edges of the surfaces.
The minimal surfaces may also have self-intersections and be non-orientable.
Finally, the minimal surfaces may have a non-trivial topology, for example, they may have handles,
see Fig. \ref{f7} (3). The
least area minimal surface for a given boundary contour is not necessary
a surface with a simplest topology.
An interesting and a separate issue is the behavior (stability) of a minimal surface against small
deformations.

Except soap films there are many other examples of minimal surfaces in Nature and human activities,
e.g in the architecture. A DNA molecule, for instance, has a structure of double helix.
The corresponding minimal surface is a helicoid,
one of the four known minimal surfaces that can be embedded in a flat 3-dimensional space
without self-intersections.

\section{Suggestion about entanglement entropy in static space-times}
\setcounter{equation}0

The surface $\cal B$ in
the entropy formula (\ref{i.1}) is embedded in a Riemannian
manifold $\bar{\cal M}$. We suppose that $\bar{\cal M}$ is a Euclidean section
of some stationary Lorentzian space-time $\bar{\cal M}^L$.
The metrics of the two spaces are related by the Wick rotation, i.e.
by the complexification of the Killing
time. If the metric is stationary but not static the Wick rotation
is accompanied by an analytical continuation
of some other parameters. For example, in an axially-symmetric space-time one
goes to complex values of the angular velocity parameter. The same change of parameters
has to be done in (\ref{i.1}).

After
the Wick rotation minimal co-dimension 2 hypersurfaces
in $\bar{\cal M}$ correspond to extremal hypersurfaces in $\bar{\cal M}^L$.
In static Lorentzian space-times extremal hypersurfaces  which end on
a given slice $\cal S$ of the
boundary have an important geometrical property.
They are minimal hypersurfaces located in a single constant
time slice $\Sigma_t$ such that $\partial \Sigma_t={\cal S}$.
To prove this statement
note that variations of the volume functional (\ref{2.11})
in $\bar{\cal M}^L$ can be written as
$\delta_X{\cal A}=\tilde{\delta}_X{\cal A}+\delta_t{\cal A}$ where $\tilde{\delta}_X{\cal A}$
are caused by shifts of the surface along $\Sigma_t$. Variations
$\delta_t{\cal A}$ are generated by shifts $\delta t$ along the time coordinate,
$\delta t=f(y)$ where $f(y)$ is an arbitrary function of coordinates on $\cal B$.
If $\bar{\cal B}$ is minimal in $\Sigma_t$ the variation of the volume becomes
\begin{equation}\label{2.11b}
\delta_X{\cal A}(\bar{\cal B})=\delta_t{\cal A}(\bar{\cal B})=
\int_{\bar{\cal B}} d^ny \sqrt{\gamma}
\gamma^{ij}X^\mu_{~,i} X^\nu_{~,i}~f_{,\mu} \xi_\nu ~~~.
\end{equation}
We noticed here that $\delta_t X^\mu=f(y)\xi^\mu$
and used the Killing identity $\xi_{\mu;\nu}+\xi_{\nu;\mu}=0$. One concludes
that a surface minimal in
$\Sigma_t$ is extremal in $\bar{\cal M}^L$ if it is orthogonal to the Killing
vector $\xi$. In static space-times the slice $\Sigma_t$ is
everywhere orthogonal to $\xi$.
We come to the following statement based on the line of reasonings of
Section 2.

\bigskip

\underline{Suggestion 1.} {\it Suppose that: 1) $\bar{\cal M}^L$ is a
static Lorentzian space-time which
is a solution to the Einstein equations; 2) constant Killing time $t$ hypersurfaces, $\Sigma_t$,
are the Cauchy hypersurfaces; 3) the boundary $\cal S$ of $\Sigma_t$
is simply connected. Consider a division of $\Sigma_t$ into two regions with boundaries $D_1$ and $D_2$
such that $D_1\bigcup D_2={\cal S}$.
Suppose that ${\cal B}$ is a minimal least volume codimension 1 hypersurface of $\Sigma_t$
which ends at $\cal S$ on
the boundary of regions
$D_1$ and $D_2$.

Under these conditions the entanglement entropy in a quantum gravity
theory associated to the partition of the system into the regions with
external boundaries $D_1$ and $D_2$ is given
by formula (\ref{i.1}) where $\cal A$ is the volume of ${\cal B}$.
The formula holds in the
leading order approximation when contributions of quantum fields
are neglected.}

\bigskip

By the definition (see \cite{HaEl:73}) any non-space-like curve
intersects the Cauchy hypersurface exactly once. The second assumption is needed
because states in a quantum theory on a curved space-time are to be defined on
a Cauchy hypersurface.

A possible extension of this suggestion to stationary but not static space-times
will be discussed in section 8 after we examine additional requirements for the entropy
formula.

\section{Black holes}
\subsection{Cauchy surfaces with disconnected boundaries}
\setcounter{equation}0

Entanglement entropy in a quantum gravity (\ref{i.1}) has the same form as the Bekenstein-Hawking
entropy of a black hole.
There are a number of arguments which indicate that entropy of black holes fits
our approach as a particular case. The black hole geometries require a special consideration
because they violate the condition  that boundary of the Cauchy surface is connected.

\begin{figure}[h]
\begin{center}
\includegraphics[height=6.3cm,width=12.0cm]{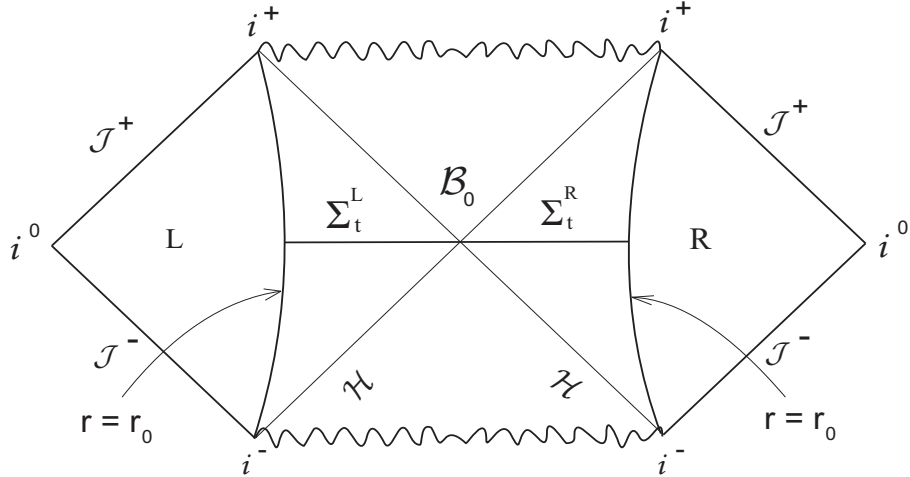}
\caption{\small{The Carter-Penrose diagram for a Schwarzschild black hole. The black hole
in cavity corresponds to a part of the diagram between the lines $r=r_0$. Shown at the diagram
is one of the Cauchy hypersurfaces $\Sigma_t^L\bigcup \Sigma_t^R$.}}
\label{f9}
\end{center}
\end{figure}

To simplify the analysis we consider only the Schwarzcshild solution
\begin{equation}\label{3.5}
ds^2=-B(r)dt^2+{dr^2 \over B(r)} +r^2d\Omega^2~~,
\end{equation}
where $B(r)=1-r_H/r$ and $r>r_h$. The metric describes the black hole geometry
outside the horizon. The horizon is located at $r=r_H$. It is
convenient to assume that the black hole is placed in spherical cavity
with the radius $r=r_0>r_H$.

The Schwarzcshild coordinates (\ref{3.5}) are not complete. The entire geometry
of an eternal black hole is shown on the Carter-Penrose diagram, Fig. \ref{f9}.
As is well known
this geometry consists of two identical, left $L$ and right
$R$, regions separated by the horizons $\cal H$. The horizons intersect at
a 2 sphere which is a bifurcation surface
${\cal B}_0$.
 Any space-like Cauchy
hypersurface in this space-time stretches from the left world to the right and ends
on the lines $r=r_0$ or, in general, connects two asymptotically flat
regions. All Cauchy hypersurfaces have two disconnected boundaries.
The black hole geometry in the each region is static.
A constant time Cauchy hypersurface $t=const$ is called the Einstein-Rosen bridge.
One of such surfaces is shown in Fig. \ref{f9}. It consists of two
parts, $\Sigma_t^L$ and $\Sigma_t^R$, which join at ${\cal B}_0$.
The Einstein-Rosen bridge has a wormhole topology.
The boundary of the bridge is $S^2\bigcup S^2$.
We denote left and right pieces of the boundary as ${\cal S}^L$ and ${\cal S}^R$,
respectively.

As before, suppose that the system is divided into two parts with external boundaries
$D_1$ and $D_2$ separated by a surface $\cal B$. A new possibility
appears in case of black holes when $D_1$ and $D_2$ do not have common points,
i.e. $D_1={\cal S}^R$ and $D_2={\cal S}^L$. In this case
the separating surface has to be
a closed surface inside the Cauchy surface. We present the arguments that
${\cal B}={\cal B}_0$.

\begin{figure}[h]
\begin{center}
\includegraphics[height=5cm,width=9.0cm]{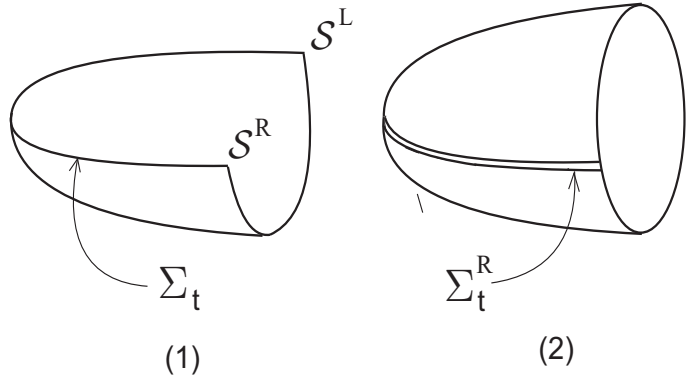}
\caption{\small{demonstrates the form of trajectories in the path integral representation for the
wave function $\Psi$ of the black hole and the density matrix
$\hat{\rho}$, figures (1) and (2), respectively.
Fig. (1) shows half of the Gibbons-Hawking instanton; (2) shows the instanton with a cut.}}
\label{f10}
\end{center}
\end{figure}

The black hole in a cavity can be in a thermal equilibrium with its radiation
\cite{York:86}.
Such a canonical ensemble is described by the partition function (\ref{2.1}) with
the boundary condition ${\cal T}=S^2\times S^1$. The length of the circle $S^1$ fixes the
temperature of the ensemble, the size of the cavity is fixed by the radius of $S^2$.
The semiclassical approximation to (\ref{2.1}) is dominated by the contribution of the
Gibbons-Hawking instanton \cite{Hawk:79}
\begin{equation}\label{3.6}
ds^2=B(r)d\tau^2+{dr^2 \over B(r)} +r^2d\Omega^2~~,
\end{equation}
where $\tau$ is periodic coordinate, $0\leq \tau \leq \beta_H$, and $\beta_H=4\pi r_H$
is the inverse Hawking temperature. The temperature of the ensemble is
$T=1/(\beta_H\sqrt{B(r_0)})$.

The black hole is
in a pure state which is called the Hartle-Hawking vacuum. A thermal nature of states
in $\Sigma_t^R$ appears when the region $\Sigma_t^L$ is unobservable.
Such point of view  was suggested long ago \cite{Israel:76}
in the framework of a thermo-field dynamics.

A more general concept, a wave-function of a
black hole, $\Psi(\varphi)$, was introduced in \cite{BFZ:95}.
Semiclassically $\Psi(\varphi)$ generates the Hartle-Hawking vacuum.
The arguments $\varphi$ of $\Psi$
(a configuration space of the theory) are 3-geometries, $\Sigma$,
and matter fields
on $\Sigma$. Spaces  $\Sigma$ have the wormhole topology $R\times S^2$. Their boundary
consists of two disconnected pieces,
${\cal S}^L$ and ${\cal S}^R$, or two asymptotically flat regions, in general.
$\Psi(\varphi)$ is defined by a path integral like (\ref{2.1}). The trajectories here are
compact 4-geometries with the boundary $\Sigma \bigcup (S^2 \times I)$ where $I$ is an interval
of the length $1/(2T)$.
An example of such a geometry is shown on Fig. \ref{f10} (1).
It is a half of the Gibbons-Hawking instanton obtained by
cutting the instanton along the bridge ($\tau=0$ and $\tau=\beta_H/2$).

Consider  a Gibbons-Hawking instanton
with a cut along  $\Sigma_t^R$, Fig. \ref{f10} (2). Let us take $n$ identical copies
of the given space and glue them along the cuts. We obtain a 4-geometry,
$\bar{\cal M}_n$, which
has conical singularities
located at the horizon $r=r_h$.
Locally $\bar{\cal M}_n$ looks as (\ref{3.6}) but has the periodicity
$0\leq \tau \leq n\beta_H$.
The geometry above can be used to compute the entanglement entropy
of field excitations located on $\Sigma_t^R$ with the help of formula (\ref{2.5}).
To this aim one has to use the partition function (\ref{2.4}) on $\bar{\cal M}_n$.
The instanton with a cut corresponds to the density matrix $\hat{\rho}$ of fields in
$\Sigma_t^R$.

To compute the entanglement entropy
in a quantum gravity theory we, like the authors of  \cite{BFZ:95},
allow quantum fluctuations of the geometry. Both the Gibbons-Hawking instanton and
position of the separating surface are to appear semiclassically when
one uses the approach of section 2 with boundary conditions adopted
to black holes.

We suggest to this aim that the entanglement entropy
for the region with the boundary $D_1={\cal S}^R$ and a black hole
inside is determined by a partition function
${\cal Z}(\beta,D_1,T)$, see  (\ref{2.51}). The partition function is represented
in terms of the integrals (\ref{2.1a}), (\ref{2.1aa}). The path integral
(\ref{2.1aa}) it is taken over a class of Riemannian 4-geometries,
$\{{\cal M}_n\}_{\cal B}$  which have the topology of $\bar{\cal M}_n$.
The boundary of these spaces coincides with the boundary of  $\bar{\cal M}_n$.
The class $\{{\cal M}_n\}_{\cal B}$ is specified by the presence of an internal
co-dimension 2
hypersurface $\cal B$ where the spaces have conical singularities ${\cal C}_\beta \times {\cal B}$,
$\beta=2\pi n$.

In the semiclassical approximation the integral (\ref{2.1aa}) is determined by a trajectory
which locally has the same geometry as (\ref{3.6}). The sum (\ref{2.1a})
picks up a class where the surface $\cal B$ has a least area on (\ref{3.6}).
Because (\ref{3.6}) is a  static space it is enough to find a least area surface
$\bar{\cal B}$ lying
in the Einstein-Rosen bridge. The spherically symmetry of $\Sigma_t$ implies that
$\bar{\cal B}$ has a constant
radius $r$ of the minimal value $r=r_H$.
Thus, the entanglement entropy is given by (\ref{i.1}) and coincides with the
Bekenstein-Hawking
entropy in accord with the interpretation of \cite{Sr:93}--\cite{FrNo:93}\footnote{A
similar interpretation of the Bekenstein-Hawking entropy on the base of minimal
surfaces  is given in \cite{Em:06} for black holes on a brane.
Different aspects of entanglement entropy as a source of black hole entropy
were discussed recently in \cite{Sol:06}--\cite{ANT:07}.}.

The surface $r=r_H$ also solves equations (\ref{3.3}). The Killing field $\partial_\tau$
has fixed points at $r=r_H$ and, as a result of this symmetry
extrinsic curvatures of $\bar{\cal B}$ vanish. Therefore,  $\bar{\cal B}$
is minimal according to (\ref{3.4}).
The same property holds for the bifurcation surface of the horizon of rotating black holes.

A spherical form of a minimal surface in a black hole space-time is supported by
the strong gravitational field near the horizon.
Closed minimal surfaces with the topology of a sphere cannot exist in
a flat space. An analog of a closed
minimal surface in a 3D flat space is a soap bubble. The bubble has a
constant non-vanishing mean curvature whose value is determined
by the surface tension and by the difference of pressures inside and outside.

\subsection{Entanglement in the presence of a black hole}

\begin{figure}[h]
\begin{center}
\includegraphics[height=12.0cm,width=14.0cm]{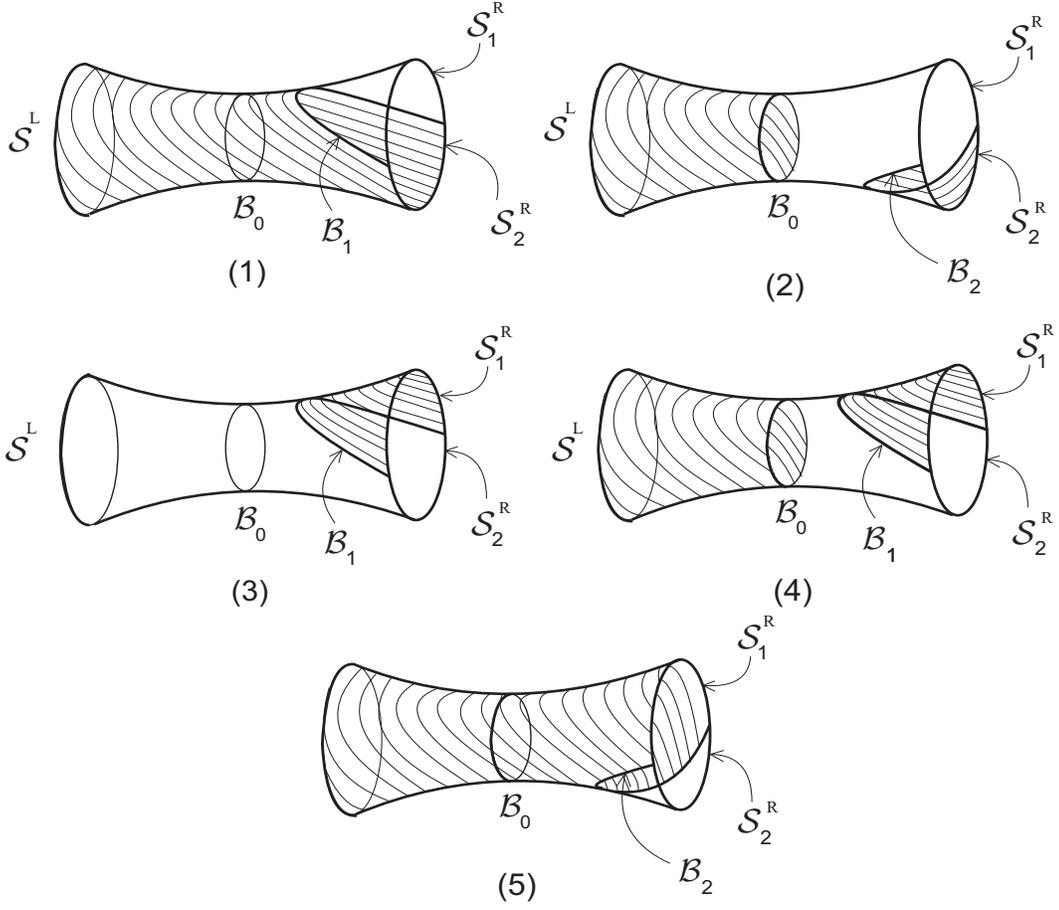}
\caption{\small{Shown
by dashed regions are
different locations
of states on the Einstein-Rosen bridge which are traced out. The boundary of these regions is
$D_2$, the boundary
of their completion is $D_1$. The cases (1),(2) correspond to the choice
$D_1=S_1^R$, $D_2=S^L\bigcup S_2^R$, the case (3) is
$D_1=S^L\bigcup S_2^R$, $D_2=S_1^R$, and the cases (4),(5) are
$D_1=S_2^R$, $D_2=S^L\bigcup S_1^R$.}}
\label{f11}
\end{center}
\end{figure}

In case of black holes the choice of the region where the states of the system are traced
out can be more non-trivial and it may not coincide with the division of the system by the horizon.
Moreover, the separating surface may consist of several disconnected elements
including the black hole horizon as one of them.
Several possibilities are shown in Fig. \ref{f11}. We suppose here that the boundary ${\cal S}^R$
of the right half of the Einstein-Rosen bridge is divided into two parts, ${\cal S}^R_1$ and
${\cal S}^R_2$.  (The left part of boundary is ${\cal S}^L$.)
A least area surface inside the bridge starts
at the boundary between ${\cal S}^R_1$ and ${\cal S}^R_2$. Locations of this surface
are denoted as ${\cal B}_1$ or ${\cal B}_2$, depending on its position with respect
to the horizon ${\cal B}_0$. Arguments given in section 2.3 indicate that among different
options one has to choose the one with the least total area.

In quantum gravity the entanglement is determined by the division
of the boundary of the system into parts  $D_1$ and $D_2$.
As earlier, we assume that the degrees of freedom in a region with the boundary $D_2$
are integrated out.  Dashed parts on Fig. \ref{f11} show
locations of this region on the Einstein-Rosen bridge for different choices of the boundary.

In cases (1),(2)  the boundary is $D_1=S_1^R$, $D_2=S^L\bigcup S_2^R$.
The first possibility, (1), is realized
when ${\cal A}({\cal B}_1)< {\cal A}({\cal B}_2)+{\cal A}({\cal B}_0)$,
the second, (2), in the opposite case. The third option is $D_1=S^L\bigcup S_2^R$, $D_2=S_1^R$,
Fig. \ref{f11}, (3). It is dual to the case (1) and is realized when
states can be measured both on the right and on the left parts of the bridge,
see Fig. \ref{f9}.
The options (1) and (3) have the same entanglement entropy, ${\cal A}({\cal B}_1)/(4G)$,
because the system is in a pure state, the Hartle-Hawking vacuum.

When states inside the black hole cannot be observed,  instead of (3) one has to consider
another definition of the boundary. It is
$D_1=S_2^R$, $D_2=S^L\bigcup S_1^R$, see Fig. \ref{f11}, (4),(5).
The corresponding entropy is either $({\cal A}({\cal B}_1)+{\cal A}({\cal B}_0))/(4G)$ or
${\cal A}({\cal B}_2)/(4G)$. In the latter case ${\cal A}({\cal B}_2)<
{\cal A}({\cal B}_1)+{\cal A}({\cal B}_0)$.
In both situations the entropy differs from the entropy in (1) because observers
outside the horizon perceive the black hole in a thermal state.
We come to the following statement regarding entanglement entropy in the presence of
black holes.

\begin{figure}[h]
\begin{center}
\includegraphics[height=4.7cm,width=6.0cm]{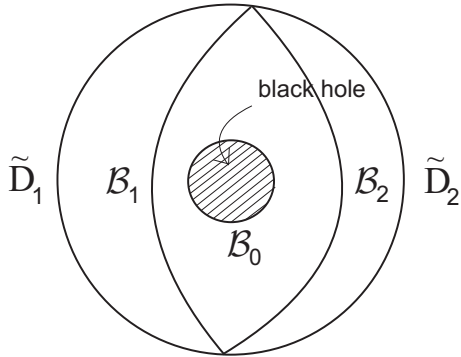}
\caption{\small{Separating surfaces in the presence of a black hole.}}
\label{f5}
\end{center}
\end{figure}

\underline{Suggestion 2:} {\it suppose that: 1) $\bar{\cal M}^L$ is a Lorentzian space-time which
is a static black hole solution to the Einstein equations;
2) $\Sigma_t$ is a part of a constant-Killing-time hypersurface in $\bar{\cal M}^L$ restricted
by the black hole horizon ${\cal B}_0$ and by an external boundary ${\cal S}$.
Consider a division of $\Sigma_t$ into two regions with
boundaries $\tilde{D}_1$ and $\tilde{D}_2$ such that $\tilde{D}_1\bigcup \tilde{D}_2={\cal S}$.
Let ${\cal B}_1$ and ${\cal B}_2$ be  codimension 1 hypersurfaces in
$\Sigma_t$ which end on the boundary between $\tilde{D}_1$ and $\tilde{D}_2$
and have least volumes among the hypersurfaces homologous to
$\tilde{D}_1$ or $\tilde{D}_2$, respectively, see Fig. \ref{f5}.

The entanglement entropy in a quantum gravity
theory associated to the loss of the information about the states
in the region with boundary $\tilde{D}_2$ is}
\begin{equation}\label{2.12-bh-1}
S(\tilde{D}_1,T)={{\cal A}({\cal B}_1) \over 4G}~~~,
\end{equation}
{\it if ${\cal A}({\cal B}_1)< {\cal A}({\cal B}_2)+{\cal A}({\cal B}_0)$, and}
\begin{equation}\label{2.12-bh-2}
S(\tilde{D}_1,T)={{\cal A}({\cal B}_2) \over 4G}+{{\cal A}({\cal B}_0) \over 4G}~~~
\end{equation}
{\it  in the opposite case. The formulae hold in the
leading order approximation when contributions of quantum fields
are neglected. If $\tilde{D}_2=\emptyset$ the entropy is the
Bekenstein-Hawking entropy of the black hole.}
Parameter $T$ in (\ref{2.12-bh-1}), (\ref{2.12-bh-2})
is a temperature of the black hole.

\section{Entropy inequalities}
\setcounter{equation}0

The suggestions made in Sections 2 and 4 cannot be rigorously proved in an effective gravity
theory because a real microscopical content of the underlying theory is not known.
The situation
is similar to the problem of the Bekenstein-Hawking  entropy $S^{BH}$.
The origin  of the entropy  of a black hole
can be understood only in a genuine quantum gravity theory.
The fact that $S^{BH}$ has properties of a physical entropy follows from different thought
experiments
which show, for example, that a generalized version of the second law of thermodynamics
holds for black holes.
In a similar way, one can check that suggestions 1 and 2 for the entanglement
entropy are physically and mathematically
consistent. The first step is to see whether they agree with basic
properties of the von Neumann entropy.

In the context of the holographic representation of the
entanglement entropy in theories with AdS duals, it was proposed \cite{HiTa:06}
to use the subadditivity property as a test for definition of the entropy.
Later Headrick and
Takayanagi \cite{HeTa:07} gave a simple geometrical proof that the
entropy in such theories possesses this property. The proof
of \cite{HeTa:07} was based on the relation between the entropy and minimal surfaces.
That is why it is also applicable under our conditions.

\begin{figure}[h]
\begin{center}
\includegraphics[height=4.8cm,width=8.5cm]{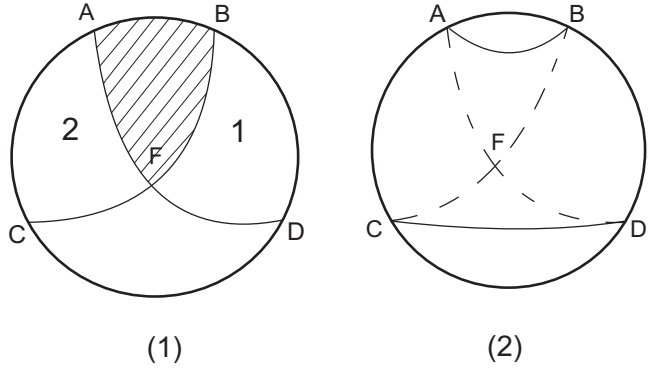}
\caption{\small{Fig. (1) demonstrates  a system with intersecting subsets (regions), 1 and 2.
Boundaries of the regions are shown
as lines $(A,D)$ and $(B,C)$. The domain of the intersection is dashed. The minimal
surfaces are the internal lines connecting pairs of points, $A$ with $D$ and $B$ with $C$.
Another set of minimal surfaces for the same configuration
of the boundaries are the
solid internal lines on Fig. (2). They connect $A$ with $B$ and $C$ with $D$.}}
\label{f6}
\end{center}
\end{figure}

Consider a set of states, $a$, in a Hilbert space of some quantum system.
We assume that $a$ corresponds to states located in
some region of the coordinate space. Let us define the reduced density matrix
$\hat{\rho}_a=\mbox{Tr}_{\bar a}\hat{\rho}$
where $\hat{\rho}$ is the density matrix of the entire system. The trace
$\mbox{Tr}_{\bar a}$  is taken over the Hilbert space except
elements $a$. The
entanglement entropy related to the given set is $S_a=-\mbox{Tr}~\hat{\rho}_a\ln \hat{\rho}_a$.
For the entropies of two sets shown on Fig. \ref{f6}, (1) one can write the following inequalities:
\begin{equation}\label{4.1}
S_1+S_2\geq S_{1\bigcup 2}+S_{1\bigcap 2}~~,
\end{equation}
\begin{equation}\label{4.2}
|S_1-S_2| \leq S_{1\bigcup 2}~~,
\end{equation}
where $S_{1\bigcup 2}$ and $S_{1\bigcap 2}$ are the entropies for the union or
intersection of the sets 1 and 2.
Formulae above are  based on the concavity of the
function $-x\ln x$ which is present in the von Neumann
definition of the entropy \cite{NiCh:00}. Equation (\ref{4.1}) is known as
the strong subadditivity property.
The Araki-Lieb inequality (\ref{4.2}) holds when the Hilbert space of the system
consists only of the two components, 1 and 2.

The sketch of the proof of (\ref{4.1}) is the following\footnote{For simplicity in this section
we put  $4G=1$.}.
If the system does not have a black hole
inside then entropies $S_1$ and $S_2$ are given by the areas of least area surfaces shown
on Fig. \ref{f6}, (1), as
internal lines connecting, respectively, pairs of points, $A$ with $D$ and $C$ with $B$.
The entropies $S_{1\bigcup 2}$ and $S_{1\bigcap 2}$ are given by least area surfaces shown
as internal lines connecting, respectively, $C$ with $D$ and $A$ with $B$,
see Fig. \ref{f6}, (2). Let us denote as ${\cal A}_{XY}$ the area of a least minimal surface
corresponding
to an internal line on Fig. \ref{f6} connecting points $X$ and $Y$.
Then
$$
S_1+S_2={\cal A}_{AD}+{\cal A}_{BC}=
{\cal A}_{AF}+{\cal A}_{FD}+ {\cal A}_{BF}+{\cal A}_{FC}=
$$
\begin{equation}\label{4.3}
({\cal A}_{AF}+{\cal A}_{BF})+({\cal A}_{FD}+{\cal A}_{FC})
\geq
{\cal A}_{AB}+{\cal A}_{DC}=
S_{1\bigcup 2}+S_{1\bigcap 2}~~~.
\end{equation}

Let us discuss now the Araki-Lieb inequality (\ref{4.2}). If the state
of the system is pure,  $S_{1\bigcup 2}=0$. In this
case (\ref{4.2}) is saturated because of the known symmetry property, $S_1=S_2$,
of the entanglement
entropy in pure states. This property is trivially satisfied
if there are no black holes:
the least area hypersurface  separating the system is uniquely determined
because the boundary of the Cauchy surface is simply connected.

In case of black holes the situation is more complicated because there may be two types
of the least area surfaces homologous to different parts of the boundary, see Fig. \ref{f11}.
Black hole is perceived by an external observer as a mixed state whose entropy
is given by the Bekenstein-Hawking formula.
Thus, (\ref{4.2}) should be written as
\begin{equation}\label{4.2a}
|S_1-S_2| \leq  S=S_0~~,
\end{equation}
where we used notations $S_1=S(\tilde{D}_1,T)$, $S_2=S(\tilde{D}_2,T)$, $S_0={\cal A}({\cal B}_0)$.
Let us show that (\ref{4.2a}) follows from the
suggestion 2.
Denote ${\cal A}_k={\cal A}({\cal B}_k)$, where $k=0,1,2$.
Suppose that ${\cal A}_2>{\cal A}_0$, and consider different situations.
If ${\cal A}_1<{\cal A}_2-{\cal A}_0$ one easily concludes that $S_1={\cal A}_1$, $S_2=
{\cal A}_1+{\cal A}_0$, and the Araki-Lieb inequality is saturated.
The same happens when ${\cal A}_1>{\cal A}_2+{\cal A}_0$ because in this case
$S_1={\cal A}_2+{\cal A}_0$, $S_2={\cal A}_2$. Finally, if
${\cal A}_2-{\cal A}_0<{\cal A}_1<{\cal A}_2+{\cal A}_0$ the entropies are
$S_1={\cal A}_1$, $S_2={\cal A}_2$. Then
$$
S_2-S_1={\cal A}_2-{\cal A}_1 < {\cal A}_0=S_0~~\mbox{for}~~S_2>S_1~~~,
$$
$$
S_1-S_2={\cal A}_1-{\cal A}_2 < {\cal A}_0=S_0~~\mbox{for}~~S_1>S_2~~~.
$$
The case ${\cal A}_2<{\cal A}_0$ is treated analogously.

Extension of these results to stationary but not static space-times will be discussed
in section 8.

\section{Variational formulae}
\setcounter{equation}0

In general, in many-body systems the entanglement entropy
cannot be expressed solely in terms of macroscopical parameters. Additional information about
the microscopical structure is needed.
That is why, in contrast with the first law
for the thermodynamical entropy, there are no universal variational formulae for
the entanglement entropy of ordinary quantum systems.
According to our suggestions the entanglement entropy $S$ in a quantum gravity
is a geometrical quantity.

We give examples how to express changes of $S$ solely in terms
of macroscopical parameters. In Minkowsky space the minimal
surface which divides the space into two parts is a plane. If one brings in
a point-like particle with mass $M$ the shape of the surface slightly changes
under the influence of the gravitational field of the particle. The space-time
metric in the weak field approximation is
\begin{equation}\label{5.1}
ds^2=-\left(1-{2MG \over R}\right)dt^2+\left(1+{2MG \over R}\right)(dx^2+dy^2+dz^2)~~~,
\end{equation}
where $R=\sqrt{x^2+y^2+z^2}$ and the particle is located at $R=0$.
It is
not difficult to see that the area of the minimal surface in the weak field approximation
is given by the integral
\begin{equation}\label{5.2}
{\cal A}=\int dxdy\left(1+{2MG \over \sqrt{x^2+y^2+z^2}}\right)~~~,
\end{equation}
where $z$ is a coordinate of the plane
in the absence of the particle. To cutoff the integral (\ref{5.2}) one may assume that
the surface
has a large but finite size. The variation of the area of the surface under the change of its
position with respect to the particle  does not depend on the cutoff
\begin{equation}\label{5.3}
\delta {\cal A}=-\delta z\int dxdy{2MGz \over (x^2+y^2+z^2)^{3/2}}=
-4\pi MG \delta z~~~.
\end{equation}
The area decreases if the distance $z$ increases.
If one restores all dimensional
constants the variational formula for
the entanglement entropy takes the following form
\begin{equation}\label{5.4}
\delta S=- \pi {Mc \over \hbar}\delta z~~~.
\end{equation}
For a particle
of the mass $1~g$ the change of the entropy under
the shift $1~cm$ is an enormous quantity, $2.7 \cdot 10^{37}$. For an elementary
particle (\ref{5.4}) can be written as
\begin{equation}\label{5.4a}
\delta S=- \pi {\delta z \over \lambda}~~~,
\end{equation}
where $\lambda=\hbar/(Mc)$ is the Compton wave length of the particle. Thus, if the particle
shifts to the distance comparable to
its Compton wave length the entropy change is
of the order of unity.

Another example where the variation of the entropy
can be easily established is the space-time around a cosmic string
\begin{equation}\label{5.5}
ds^2=-dt^2+\left(1-4\mu G\right)^2\rho^2 d\varphi^2+d\rho^2+dz^2~~~,
\end{equation}
where $0\leq \varphi < 2\pi$.
The string is located at $\rho=0$ and has a tension $\mu$. The gravitational effects
around the string result in a deficit of the polar angle around  its axis.
A minimal surface which is parallel to the string is a plane slightly
distorted by  the string gravity. Suppose that
in the Cartesian coordinates ($x=\rho \cos\varphi$, $y=\rho \sin\varphi$)
the position of the plane
is $x=l$. Then in the weak field approximation, $\mu G \ll 1$,
the area of the minimal surface per unit length is
\begin{equation}\label{5.6}
a=\int dy\left(1-4\mu G {l^2 \over l^2+y^2}\right)~~.
\end{equation}
The parameter $l$ is a distance from the string to the minimal surface.
Note that the deviation from the area of the plane per unit length is finite,
$-4\pi \mu G l$. Change of the entropy (per unit length)
under the shift of the string $\delta l$ directed outward of the minimal surface is
\begin{equation}\label{5.7}
\delta s=- \pi \mu\delta l~~~.
\end{equation}
In Planck units $\mu \sim 10^{-6}$ for GUT strings ($10^{16}~GeV$) and
$\mu \sim 10^{-34}$ for electroweak strings ($100~GeV$).
One can consider deformations of the strings when a portion of the string of a size $z$
moves toward or outward of the plane to a distance $\delta l$.
Changes of
the entropy caused by such deformations can be estimated as $\delta S\sim z \delta s\sim
-\mu z \delta l$.
It is easy to see that $\delta S$ is of the order of unity when both $z$ and $\delta l$ are
determined
by the scales of the corresponding theory,
($z\sim \delta l \sim 10^{-30}~cm$ for GUT strings, or $z\sim \delta l \sim 10^{-16}~cm$
for electroweak strings). If $z$ or $\delta l$ are macroscopical scales the entropy
changes are very large.

\section{Quantum corrections}
\setcounter{equation}0

In the Gibbons-Hawking approach the low-energy fields $\phi$ yield quantum corrections
to the semiclassical approximation (\ref{2.10a}).
Quantum effects change the classical solution $\bar{g}_{\mu\nu}$
because of a stress-energy tensor of the quantum matter in the r.h.s. of
the Einstein equations (\ref{2.10}). As a result, deformations of the minimal surface $\cal B$
caused by the back-reaction change of the entanglement entropy (\ref{i.1}).

There may be another source of quantum corrections to the entropy.
Entanglement of the low-energy fields separated by the surface $\cal B$
is quantified by its own entropy. We denote this entropy $S_{q}(\cal B)$.
In an effective theory $S_{q}(\cal B)$ is an ultraviolet divergent quantity and
leading divergences of $S_q(\cal B)$ are proportional to the area
${\cal A}({\cal B})$ \cite{Sr:93}, \cite{BKLS}\footnote{For the discussion
of the divergences of the entanglement entropy for a system
with a boundary in Minkowsky space-time, see \cite{DF:06a}, \cite{Casini}
and references therein.}.
Thus, the "gravitational" entropy,
${\cal A}({\cal B})/(4G)$, and $S_{q}(\cal B)$ may be related.
We make the following statement.

\underline{Suggestion 3:} {\it In a static space-time
entanglement entropy $S_{q}(\cal B)$ of the low-energy matter fields divided
by a minimal hypersurface $\cal B$
contributes to a quantum correction to the entropy (\ref{i.1}).
The leading ultraviolet divergences of $S_{q}(\cal B)$ are removed in the course
of a standard renormalization of the gravity coupling $G$.}

The idea of renormalizing  the divergences in the
entanglement entropy was first formulated for fields
around a black hole \cite{CaWi:94},\cite{SuUg:94}
(for a review, see \cite{FrFu:98}).
We emphasize that reasonings applicable to black hole horizons
can be extended to minimal hypersurfaces in a constant-time slice.

Subleading divergences in the effective action require
terms which depend quadratically on components of the curvature tensor.
The form of the entropy (\ref{i.1}) in the presence of $R^2$-terms is modified.
Once this modification is taken into account, we suppose that subleading
divergences of $S_{q}(\cal B)$ are also eliminated in the course
of renormalization of the corresponding couplings at $R^2$-terms.

Suggestion 3 gives a prescription how to do computations which take into account
entanglement entropy of matter fields. Its consequences and applications will be
considered elsewhere.

\section{Concluding remarks}
\setcounter{equation}0

We presented a line of reasonings
that the entanglement entropy $S$ in a quantum gravity theory
is defined by macroscopical parameters of the system. Like a thermodynamical
entropy, $S$ does not require knowing
a genuine microscopical structure. In the leading order $S$ is given
by the Bekenstein-Hawking formula (\ref{i.1})
in terms of the volume of a co-dimension 2 hypersurface $\cal B$. For given boundary
conditions $\cal B$ must be a least volume hypersurface
embedded in a Euclidean section $\bar{\cal M}$
of the corresponding Lorenzian solution.
Our reasonings are based on the assumption that gravity is an emergent phenomenon
and we make use of the Gibbons-Hawking path integral as a low-energy definition of the partition
function.

In static space-times
$\cal B$ can be interpreted as  a real hypersurface
on a constant time slice. Thus, despite of fluctuations of the geometry
a spatial
division of degrees of freedom in a quantum gravity theory can have a meaning
in a semiclassical approximation
provided that the separating surface has a least volume in the slice.

The idea that the concept of a "gravitational" entropy can be extended to co-dimension 2
hypersurfaces is not new. In \cite{Jac:95} Jacobson applied  the Bekenstein-Hawking formula
to small space-like 2-surfaces in arbitrary 4D space-times.
The past directed null normal congruences of the surfaces
("local Rindler horizons") were supposed to have vanishing expansion and shear
at a point. In \cite{MP} an attempt has been made to extend this
construction to arbitrary 2-surfaces with a finite area.
What differs our approach from the above suggestions is that we directly relate
the entropy to quantum entanglement and on this base derive dynamical equations
for the surface.

In static space-times properties of the entanglement entropy $S$ are relatively simple.
Although the underlying degrees of freedom are not known
one can demonstrate here that $S$ satisfies a subadditivity property, which is
a necessary requirement. This happens because $\cal B$ is defined as a least volume
hypersurface.

The next step would be to define the entropy in case of stationary but not static space-times.
Our approach indicates that $S$ is still given by (\ref{i.1}) where $\cal B$
is a minimal least volume co-dimension
2 hypersurface on a Euclidean section $\bar{\cal M}$ of a corresponding physical
space-time $\bar{\cal M}^L$.
The physical value of the entropy should be attained after analytical continuation. After
the Wick rotation minimal co-dimension 2 hypersurfaces
in $\bar{\cal M}$ correspond to extremal hypersurfaces in $\bar{\cal M}^L$.

If $\bar{\cal M}^L$ is stationary but not static an extremal surface $\cal B$
does not lie in a constant-time slice unless $\cal B$ is orthogonal to the Killing field
generating time translations, see section 3.
Thus, in general different surfaces $\cal B$ may separate the system in different space-like slices.

One can still try to give a definition of the entropy which
is consistent with the basic features such as the subbadditivity property.
It is natural to guess that in stationary space-times
$\cal B$  in (\ref{i.1}) should be an extremal hypersurface
with a least volume for given boundary conditions. In certain examples such surfaces
can be obtained from the corresponding least volume surfaces in Euclidean theory.
If this is the case one can prove the
subbaditivity property along the lines of section 5. The proof in section 5
does not require the surfaces to be in a single slice.

Generalization of our construction to time-dependent geometries should be the
subject of a separate discussion. Some proposals in the context of the holographic
formula for the entanglement entropy in conformal theories are discussed in \cite{HRT}.

\bigskip

\noindent
\section*{Acknowledgment}\noindent
I am grateful N.A. Tyurin and A.I. Zelnikov for very helpful discussions.
The author thanks the Asia Pacific Centre for Theoretical Physics for the support
during the Focus Program on New Frontiers in Black Hole Physics
(Pohang, 2007).
This work was also supported by the Scientific School Grant N 5332.2006.2.

\end{document}